 \documentclass[iop]{emulateapj}

\usepackage{graphicx,natbib}
\usepackage{hyperref}

\slugcomment{Accepted to the Astronomical Journal}

\shorttitle{SN site IFU spectroscopy -- Ib/c progenitors}
\shortauthors{Kuncarayakti et al.}

\begin{document}


\title{Integral field spectroscopy of supernova explosion sites: constraining mass and metallicity of the progenitors - I. Type Ib and Ic supernovae}


\author{Hanindyo Kuncarayakti}
\affil{Kavli Institute for the Physics and Mathematics of the Universe, the University of Tokyo, 5-1-5 Kashiwanoha, Kashiwa, Chiba 277-8583, Japan}
\affil{Institute of Astronomy, Graduate School of Science, the University of Tokyo, 2-21-1 Osawa, Mitaka, Tokyo 181-0015, Japan}
\affil{Department of Astronomy, Graduate School of Science, the University of Tokyo, 7-3-1 Hongo, Bunkyo-ku, Tokyo 113-0033, Japan}
\email{hanindyo.kuncarayakti@ipmu.jp}

\author{Mamoru Doi}
\affil{Institute of Astronomy, Graduate School of Science, the University of Tokyo, 2-21-1 Osawa, Mitaka, Tokyo 181-0015, Japan}
\affil{Research Center for the Early Universe, the University of Tokyo, 7-3-1 Hongo, Bunkyo-ku, Tokyo 113-0033, Japan}

\author{Greg Aldering}
\affil{Physics Division, Lawrence Berkeley National Laboratory, 1 Cyclotron Road, Berkeley, CA 94720}

\author{Nobuo Arimoto}
\affil{National Astronomical Observatory of Japan, 2-21-1 Osawa, Mitaka, Tokyo 181-0015, Japan}
\affil{Subaru Telescope, National Astronomical Observatory of Japan, 650 North A\rq{ohoku} Place, Hilo, HI 96720}

\author{Keiichi Maeda}
\affil{Kavli Institute for the Physics and Mathematics of the Universe, the University of Tokyo, 5-1-5 Kashiwanoha, Kashiwa, Chiba 277-8583, Japan}

\author{Tomoki Morokuma}
\affil{Institute of Astronomy, Graduate School of Science, the University of Tokyo, 2-21-1 Osawa, Mitaka, Tokyo 181-0015, Japan}

\author{Rui Pereira}
\affil{CNRS/IN2P3, Institut de Physique Nucl\'eaire de Lyon, 4 Rue Enrico Fermi, 69622 Villeurbanne Cedex, France}

\author{Tomonori Usuda}
\affil{Subaru Telescope, National Astronomical Observatory of Japan, 650 North A\rq{ohoku} Place, Hilo, HI 96720}

\author{Yasuhito Hashiba}
\affil{Institute of Astronomy, Graduate School of Science, the University of Tokyo, 2-21-1 Osawa, Mitaka, Tokyo 181-0015, Japan}
\affil{Department of Astronomy, Graduate School of Science, the University of Tokyo, 7-3-1 Hongo, Bunkyo-ku, Tokyo 113-0033, Japan}





\begin{abstract}
Integral field spectroscopy of 11 type-Ib/c supernova explosion sites in nearby galaxies has been obtained using UH88/SNIFS and Gemini-N/GMOS. The use of integral field spectroscopy enables us to obtain both spatial and spectral information of the explosion site, allowing the identification of the parent stellar population of the supernova progenitor star. The spectrum of the parent population provides metallicity determination via strong-line method and age estimation obtained via comparison with simple stellar population (SSP) models. We adopt this information as the metallicity and age of the supernova progenitor, under the assumption that it was coeval with the parent stellar population. The age of the star corresponds to its lifetime, which in turn gives the estimate of its initial mass. With this method we were able to determine both the metallicity and initial (ZAMS) mass of the progenitor stars of the type Ib and Ic supernovae. We found that on average SN Ic explosion sites are more metal-rich and younger than SN Ib sites. The initial mass of the progenitors derived from parent stellar population age suggests that SN Ic have more massive progenitors than SN Ib. In addition, we also found indication that some of our SN progenitors are less massive than $\sim25$~M$_\odot$, indicating that they may have been stars in a close binary system that have lost their outer envelope via binary interactions to produce Ib/c supernovae, instead of single Wolf-Rayet stars. These findings support the current suggestions that both binary and single progenitor channels are in effect in producing type Ib/c supernovae. This work also demonstrates the power of integral field spectroscopy in investigating supernova environments and active star forming regions.
\end{abstract}





\section{Introduction}
A great number of supernovae (SNe) have been observed and new ones are being discovered owing to ongoing transient surveys and monitorings. Despite the large, growing number of observed events the real nature of the progenitor of each type of supernova is still not very well understood. 

Recently, a number of core-collapse supernovae (CCSNe) have had their progenitor star detected in high resolution pre-explosion images (e.g. \citet{vandyk12,fraser11,eliasrosa11}; see \citet{smartt09araa} for a review). The detection of the progenitor in several passbands allows the comparison between observed colors and luminosity of the star with stellar models. While direct detection is obviously desirable, in most cases only upper luminosity limits could be derived due to non-detection. This, however, is still useful in constraining the progenitor properties. Comparison with theoretical models yields the evolutionary stage and position in the Hertzprung-Russel diagram of the star, from which the initial mass could ultimately be derived. Using this method \citet{smartt09} concludes that SNe II-P arise from red supergiant stars with initial mass between $8.5^{+1}_{-1.5}$ and $16.5\pm1.5$~M$_\odot$. 

While the progenitors of SNe II-P have been convincingly detected and characterized, currently no Ib/c progenitor has been detected. It is widely believed that since SNe Ib/c are deficient in hydrogen and helium, the most likely candidate for their progenitors would naturally be the Wolf-Rayet (WR) stars \citep{crowther07}. Numerous stellar evolution models have been proposed to characterize the progenitor stars of different SN types \citep[e.g.][]{heger03, eldridge04, georgy09} but these predictions are still not very well tested with observational data. Progenitor discoveries could provide useful comparisons but the number of data points is still too small and not very reliable, considering a number of uncertainties such as metallicity assumption and also the sensitivity of the method to errors in luminosity (determined using host galaxy distance) and the end state of the evolution of the exploding star. Furthermore, progenitor discoveries ideally should be followed-up by post-explosion observation to ensure that the purported progenitor has indeed gone. However, this has not been done for the majority of the progenitors thus future confirmation of their disappearance is still needed.

Initial mass of a progenitor star is generally considered as the most influential parameter driving the evolution of the star. Hydrogen-deficient (type-Ib, Ic) SNe require progenitors with high mass-loss rate to remove the outer hydrogen envelope of the star, in which metallicity-driven wind \citep[e.g.][]{vink01} or other mechanism could play a significant role. Therefore, it is also important to investigate the metallicity of the progenitor star in addition to the initial mass. This dependence on metallicity affects the number ratio $N_{Ib/c}/N_{II}$ as suggested by several previous studies \citep{prantzos03,boissier09,eldridge08}. Another possible mechanism of mass loss is via close binary interaction \citep[e.g.][]{podsiadlowski92}. Extending the work of \citet{smartt09} for SN Ib/c progenitors, \citet{eldridge13} suggest that the non-detection of SN Ib/c progenitors may be explained by the idea that probably most of SN Ib/c progenitors arise from interacting binaries with mass $\lesssim20$ M$_\odot$. An example where binary scenario provides a good alternative explanation to the massive single star progenitor is given by \citet{crockett07} in the case of SN Ic 2002ap. From analysis of pre-explosion images, they suggest that the progenitor of SN 2002ap was probably initially a 30--40 M$_\odot$ star with twice the standard mass-loss rate, or in a binary system as a 15--20 M$_\odot$ star.

\citet{anderson10} showed that SNe Ib/c occur in higher metallicity environments than SNe II. High metallicity massive stars tend to explode as SNe Ib/c, thus higher number ratio $N_{Ib/c}/N_{II}$ is expected in higher-metallicity environment. While \citet{anderson10} suggest that SN Ib and Ic have similar metallicities, \citet{modjaz11} argue that SN Ic have higher metallicity than Ib. The results of \citet{leloudas11}, and recently \citet{sanders12}, also show indications for higher metallicity for Ic, but those were not statistically significant. In contrast with this work, all those previous studies mentioned above used slit spectroscopy to obtain spectra of the SN explosion sites, from which the site metallicities were derived.

Considering the limitations of finding SN progenitors in high-quality pre-explosion images, several alternative methods have been devised to obtain insights into the nature and possible mass range of the progenitors. A statistical study by \citet{aj08} using correlations of SN positions within host galaxies with H$\alpha$ emission showed that the association of SNe with recent star formation follows a SN Ic $>$ Ib $>$ II sequence, indicating a grouping of decreasing progenitor mass. This result is further reinforced in \citet{anderson12}; but see \citet{crowther13} for a discussion on how SN association with a nearby H\texttt{II} only constrains the progenitor mass weakly. \citet{leloudas11} suggest that SN Ib/c progenitors are more massive than 25--30~M$_\odot$ from the age of the youngest stellar population at the explosion site, estimated using H$\alpha$ emission line equivalent width. However, in that case it is likely that multiple stellar populations have been observed and treated as a single one since the study used integrated light over a large region within the spectrograph slit, in some cases $\sim1$ kpc. Some of their targets are also not the exact explosion sites but other regions several kpc away in the host galaxy. Therefore, consequently the real parent population might has been contaminated by other populations of possibly different ages and metallicities. This problem is also present in similar studies such as \citet{modjaz11} and \citet{sanders12}. Here we endeavour to refine the results of those efforts by resolved spectroscopy of the explosion site, with typical spatial resolution of around 70 pc (1" at 15 Mpc). A similar approach has been employed by \citet{gogarten09} to estimate the initial mass of the progenitor of the optical transient NGC 300 OT2008-1. They probed region $\sim$50 pc around the transient, and determined the initial progenitor mass of 12--25 M$_\odot$ from the analysis of the stellar population at the site.

In this paper we report the results of our study of 11 nearby Ib/c SN explosion sites, in which the parent stellar populations were detected and used to derive progenitor star\rq{}s mass and metallicity. We will use the terms star cluster, H\texttt{II} region, and OB association interchangeably in this paper to refer to the stellar population. As stars are born in clusters \citep[][also see \citet{bressert10} who suggest that this depends on the adopted cluster definition, which may change the fraction of stars born in clusters to be $\sim45$--90\%]{lada03} and there is evidence that all massive stars may be born within clustered environments \citep{pz10}, it is reasonable to derive a coeval star\rq{}s age and metallicity from its parent stellar population. It is not expected that the progenitor star would have wandered far from the birthplace cluster since they would not have much time to travel before exploding as a supernova -- thus the parent cluster may be studied to extract useful information on the coeval SN progenitor star. Typical velocity dispersion inside a star cluster is a few km/s \citep{bastian06,pz10}, which corresponds to about a few pc/Myr. Therefore a SN progenitor which belongs to such cluster, if unbound, would only travel typically a few tens of parsec during its several Myr lifetime. By studying the host cluster we were able to derive initial mass and metallicity of the SN progenitors. 

We organize the paper as follows. Observation, data reductions, and the subsequent analysis is presented in Section \ref{data}. Descriptions of each SN explosion sites are given in Section \ref{sites}. We discuss our findings in Section \ref{discu} and finally a summary is presented in Section \ref{summ}.

\bigskip
\section{Data acquisition and analysis method}
\label{data}
We selected our samples from the Asiago Supernova Database \citep{barbon99}. SNe in host galaxies with radial velocity of 3000 km/s or less were selected and subsequently careful visual inspections were performed to DSS and SDSS images of the explosion sites using ALADIN\footnote{\url{http://aladin.u-strasbg.fr/aladin.gml}}. We use equatorial coordinates from the Asiago database to locate the SN positions in the images. In addition, we also used the published images of the SN environment/light echo study by \citet{boffi99} to compare with and identify some SN host clusters. SN classification follows the Asiago database classification; in this paper we present the study of our SNe Ib/c samples while the type-II SNe are presented in \citet[Paper II]{paper2}. 

SNe showing close association with a bright knot were selected as observation targets. We interpret these knots as the parent stellar population of the SN progenitor star. The utilization of broadband images ensure that there is no selection preference towards very young stellar populations, as opposed to if we carried out our selection using H$\alpha$ or UV images. Therefore, we expect very little age bias to be present in our sample explosion site. 

We observed 7 of our targets in 2010-2011 using the SuperNova Integral Field Spectrograph (SNIFS; \citet{aldering02}, \citet{lantz04}) mounted at University of Hawaii 2.2 m telescope (UH88) at Mauna Kea. Table \ref{tabobs} contains the observations of our SN site targets. The positional uncertainty of each SN is also showed; the reasons for each estimates are given in the description of each explosion sites. 

The integral field spectrograph SNIFS employs a fully-filled $15\times15$ lenslet array in the integral field unit (IFU), covering 6.4"$\times$6.4" wide field of view (FoV). The corresponding spatial resolution is 0.43 arcsec per spatial pixel (spaxel). 
SNIFS has a photometric channel that is used to accurately place objects in the IFU. The positional accuracy with which a given (RA, Dec) coordinates is placed on the IFU is around 0.2 arcsec. The scale and rotation of the SNIFS IFU is well measured thus should not produce much additional astrometric uncertainty.
The spectrograph consists of two arms operating simultaneously to cover the whole optical wavelength range at $R\sim1000$: the blue channel covers 3300 -- 5200 \AA$ $ and the red channel covers 5100 -- 9700 \AA. SNIFS is operated remotely with a fully-dedicated data reduction pipeline. The final pipeline result is wavelength- and flux-calibrated ($x,y,\lambda$) datacubes. The description of the pipeline is similar to what is presented in \S 4 of \citet{bacon01} and is outlined briefly in \citet{aldering06}. 

Observation of the other 4 targets was carried out using the Gemini Multi-Object Spectrograph (GMOS) in IFU mode \citep{allington02,hook04}, with the 8.1 m Gemini North telescope at Mauna Kea (Program ID GN-2011B-C-6, via Subaru--Gemini Time Exchange Program). GMOS was used in one-slit mode, providing a 5" $\times$ 3.5" field of view sampled by the hexagonal lenslet array with one lenslet diameter of 0.2". We used the B600 grating centered at 5400 \AA, and also took identical exposures with the grating centered at 5450 \AA$ $ in order to perform spectral dithering to eliminate cosmic rays and cover the gaps of the three CCDs utilized by GMOS. This enables us to observe continuous spectra from 4000 \AA$ $ to 6800 \AA$ $ at $R\sim1700$, thus covering and resolving prominent spectral lines such as H$\alpha$, H$\beta$, H$\gamma$, H$\delta$, [O\texttt{III}]$\lambda\lambda$4959,5007, [N\texttt{II}]$\lambda\lambda$6548,6583, and [S\texttt{II}]$\lambda\lambda$6716,6731. Subsequent data reduction was carried out using the Gemini package in IRAF\footnote{IRAF is distributed by the National Optical Astronomy Observatories, which are operated by the Association of Universities for Research in Astronomy, Inc., under cooperative agreement with the National Science Foundation.}. The data was reduced to remove instrumental signatures and then calibrated in wavelength and flux to produce final ($x,y,\lambda$) datacubes.

Once the raw SNIFS and GMOS data have been fully reduced into ($x,y,\lambda$) datacubes, object spectrum extraction and analysis were done using IRAF. The datacubes were treated as stacks of images taken at different wavelengths. The flux density of an object at each wavelength 'slice' was extracted by performing aperture photometry (IRAF/\textit{apphot}) on the object at each slice, and arranging the values in the wavelength direction to produce a spectrum. The seeing FWHM was used as the radius of the circular aperture, or in some cases smaller apertures were used due to object proximity to the edge of the field of view. Sky in SNIFS data was estimated using an annular aperture around the object and subtracted out. Usually the annulus is larger than the field width thus effectively only a part of it was being used to measure the sky. GMOS is equipped by another IFU dedicated for simultaneous sky observation during the science exposures, and subtraction was done in the reduction process. 

The positions of the objects in each SNIFS wavelength slice were traced, thus compensating the differential atmospheric refraction effect \citep[DAR;][]{filippenko82}. This demonstrates the advantage of using integral field spectroscopy (IFS) compared to conventional slit spectroscopy which is strongly affected by DAR, and with IFS the effects of DAR could be corrected \textit{a posteriori} as has been shown by \citet{arribas99}. The GMOS dataset was corrected for DAR in the datacube stage by shifting each wavelength slices, and aperture photometry of individual slices was done with fixed aperture position for each object. 

The final resulting spectra were then analysed using IRAF/\textit{splot}. Nebular emission lines were measured by fitting a Gaussian profile. To measure metallicity, only the nebular emission lines are needed thus the stellar continuum was removed prior to line measurement by fitting a polynomial function. Before measuring line equivalent width, the continuum was normalized also by fitting with a polynomial. Metallicity in terms of oxygen abundance was derived using the O3N2 and N2 calibrations by \citet[hereafter PP04]{pp04}. This method of metallicity determination uses intensity ratios of lines closely spaced in wavelength, therefore not sensitive against errors introduced by reddening and flux calibration.

Following PP04, we adopt 12 + log (O/H) = 8.66 as the value of the solar oxygen abundance \citep{asplund04}. In all of the observed fields H$\alpha$ and [NII]$\lambda$6583 were detected thus enabling N2 determination, while the H$\beta$ and [OIII]$\lambda$5007 detection required for O3N2 is less frequent. In cases where both O3N2 and N2 were detected we adopt the mean value as the metallicity, otherwise the N2 value is adopted. The errors quoted in metallicity in Z$_\odot$ unit are the bounds for the highest and lowest values of 12+log(O/H) determined from O3N2 and N2. In cases where only N2 is available, we assign the error of 12+log(O/H) as $\pm 0.18$ dex, which is the 1$\sigma$ uncertainty in the PP04's N2 calibration determination. These were then converted into Z$_\odot$ metallicities to give the highest and lowest value bounds of metallicity.

\begin{figure}[Ht!]
\plotone{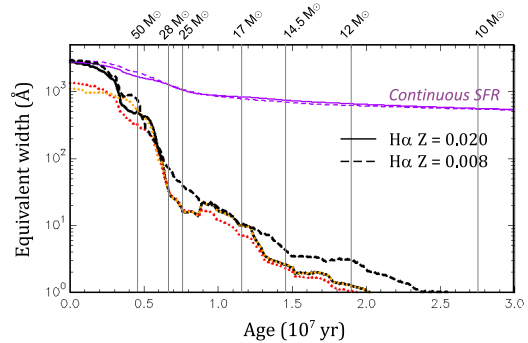}
\caption{Time evolution of H$\alpha$ equivalent width, Starburst99 SSP. Single-burst model is represented with black line color and continuous star formation with purple; both use standard Salpeter IMF with $\alpha = 2.35$, M$_{\textrm{up}} = 100$ M$_\odot$. Dotted lines represent single-burst solar-metallicity models with different IMFs: red is for $\alpha = 3.30$, M$_{\textrm{up}} = 100$ M$_\odot$, orange is for $\alpha = 2.35$, M$_{\textrm{up}} = 30$ M$_\odot$.
The lifetimes of single stars of different initial masses at solar metallicity according to Padova models are indicated with vertical grey lines.}
\label{haew}
\end{figure}

Ages of the stellar populations were determined by comparing the observed H$\alpha$ equivalent width (EW) with the theoretical values provided by simple stellar population (SSP) model Starburst99 \citep{leitherer99} for each appropriate metallicity. We assume an instantaneous-burst stellar population with continuous, standard Salpeter IMF ($\alpha=2.35$). The evolution of H$\alpha$EW with SSP age is presented in Figure~\ref{haew}. H$\alpha$EW of a stellar population with continuous star formation is also shown for comparison. The evolution of H$\alpha$ is sensitive to the star formation history, while IMF variations give less important effect especially at SSP age older than $\sim4$ Myr. The error of H$\alpha$EW were estimated from the signal-to-noise ratio of the spectral continuum around H$\alpha$. We expect the effect of dust absorption in H$\alpha$EW determination to be minimal due to the very small wavelength range used for EW measurement. The SSP age determined from H$\alpha$EW is adopted as the lifetime of the coeval progenitor star, and comparison with Padova stellar evolution models gives the estimated initial mass of a star with such lifetime. We used models from \citet{bressan93} for solar metallicity (Z = 0.02), and \citet{fagotto94} for 0.4 solar metallicity (Z = 0.008). We defined observed oxygen abundance of 0.7 (O/H)$_\odot$, corresponding to 12+log(O/H) = 8.50, as the dividing line between using the solar-metallicity model or the subsolar one to determine stellar initial mass. These models extend from the zero-age main sequence (ZAMS) to the onset of central carbon burning. This is considered representative of the stellar lifetime since the time elapsed from carbon burning to SN is only about a few hundred years, negligible compared to the $\sim10^{6}-10^7$ year lifetime of the star.

\begin{deluxetable*}{lcrrclrlccc}
\tabletypesize{\scriptsize}
\tablecaption{Target SN sites IFU observations}
\tablewidth{0pt}
\tablehead{
\colhead{SN} & \colhead{Type} & \colhead{RA2000} & \colhead{Dec2000} & \colhead{$\sigma_{\alpha,\delta}$} & \colhead{Galaxy ($N_{\textrm{SN}}$)} &
\colhead{$d$/Mpc\tablenotemark{a}} & \colhead{Obs. date\tablenotemark{b}} & \colhead{Inst.\tablenotemark{c}} & \colhead{Exposure} &
\colhead{Seeing} 
}
\startdata
1964L & Ic & 11:52:49.09  & +44:07:45.4 & $\pm2$" & NGC 3938 (3)    &   17.4 &  2011 Mar 13 & S & 1800 s $\times2$  & 0.8"     \\
1994I & Ic & 13:29:54.07 & +47:11:30.5 & $\pm1$" & NGC 5194 (4)   &   7.9 &  2011 Mar 10 & S & 1800 s $\times2$  & 1.1"     \\
2000ew & Ic & 11:40:58.52 & +11:27:55.9 & $\pm0.7$" & NGC 3810 (2) & 17.8 & 2011 Mar 15 & S & 1800 s $\times2$ & 0.9" \\
2004gt & Ic & 12:01:50.37 & $-$18:52:12.7 & $\pm0.005$" & NGC 4038 (4) & 20.9 & 2011 Mar 10 & S & 1800 s $\times2$ & 1.2" \\
2007gr & Ic & 02:43:27.98 & +37:20:44.7 & $\pm0.02$" & NGC 1058 (3) & 7.7 & 2011 Sep 29 & G & 1800 s $\times4$ & 0.5" \\
2009em & Ic & 00:34:44.53 & $-$08:23:57.6 & $\pm1$" & NGC 157 (1) & 19.5 & 2011 Sep 29 & G & 1800 s $\times2$ & 0.6" \\
1983N & Ib & 13:36:51.23 & $-$29:54:01.7 & $\pm0.6$"  & NGC 5236 (5) & 6.9 & 2011 Mar 11 & S & 1800 s $\times1$ & 1.5" \\
1984L & Ib & 02:35:30.54 & $-$07:09:30.3 & $\pm0.3$" & NGC 991 (1) & 18.8 & 2011 Sep 29 & G & 2700 s $\times2$ & 0.6" \\
1999ec & Ib & 06:16:16.16 & $-$21:22:09.8 & $\pm0.2$" & NGC 2207 (4) & 31.3 & 2011 Mar 10\tablenotemark{d} & S & 1800 s $\times2$ & 1.1" \\
2008bo & Ib & 18:19:54.41 & +74:34:21.0 & $\pm0.2$" & NGC 6643 (2) & 22.1 & 2010 Aug 1 & S & 1800 s $\times1$ & 0.8" \\
2009jf & Ib & 23:04:52.98 & +12:19:59.5 & $\pm0.06$" & NGC 7479 (2) & 33.9 & 2011 Sep 29 & G & 1800 s $\times4$ & 0.5" \\
\enddata
\tablenotetext{a}{Mean redshift-independent distance from NED (http://ned.ipac.caltech.edu).}
\tablenotetext{b}{Hawaiian Standard Time (UTC -- 10).}
\tablenotetext{c}{Instrument used: S = SNIFS, G = GMOS.}
\tablenotetext{d}{Great East Japan Earthquake occurred during the exposure of this object; observation was conducted remotely from Tokyo but fortunately the earthquake did not introduce any effect to the observation and data.}
\label{tabobs}
\end{deluxetable*}

\begin{figure}[Ht!]
\epsscale{1.}
\plotone{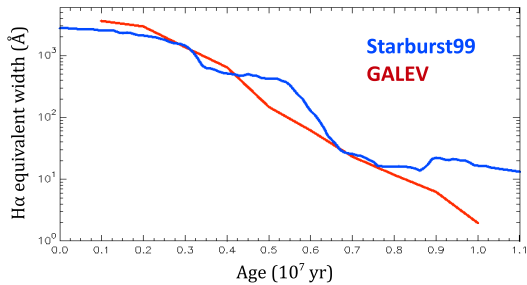}
\caption{Comparison of H$\alpha$ equivalent width evolution from different SSP models, Starburst99 and GALEV.}
\label{sspcomp}
\end{figure}

We compare Starburst99 model of H$\alpha$EW against GALEV \citep{anders03,kotulla09}, to check for consistency and possible systematic effects. Both SSP models consider the contribution of stellar and nebular continuum to the output light, but GALEV includes nebular line emissions. As for the H$\alpha$EW value as a function of SSP age, Starburst99 provides the tabulation of the values while GALEV does not. Therefore, individual spectra of GALEV SSP models of different ages were extracted and then H$\alpha$ equivalent widths were measured using IRAF/\textit{splot} with the same procedure as the one applied to observational data. The result comparing the two different SSP models is presented in Figure \ref{sspcomp}. For both Starburst99 and GALEV we assume an instantaneous burst of a stellar population with Salpeter IMF at solar metallicity, using Geneva stellar library. From the plot it is evident that H$\alpha$EW may differ by a factor two or more between the two models. We attribute this difference partly because of the resolution difference of the model SED. GALEV SEDs have resolution of 20 \AA$ $ at H$\alpha$, thus the neighboring nitrogen lines ([N\texttt{II}]$\lambda\lambda$6548,6583) were not resolved thus contaminated the H$\alpha$ in the EW measurement. However, this EW difference translates into only a small difference in age, in the order of 1--2 Myr. This, eventually, will lead into difference of typically only about 20--30 \% or lower in star initial mass, depending on the region of the age probed by H$\alpha$ equivalent width. Therefore, we conclude that the selection of SSP model does not introduce significant sytematics to the final result of progenitor star initial mass. In this case, Starburst99 was preferred over GALEV since it has better temporal resolution.

\bigskip
\section{The explosion sites}
\label{sites}

\medskip
\subsection{SNe Ic sites}

\begin{figure*}[Ht!]
\plotone{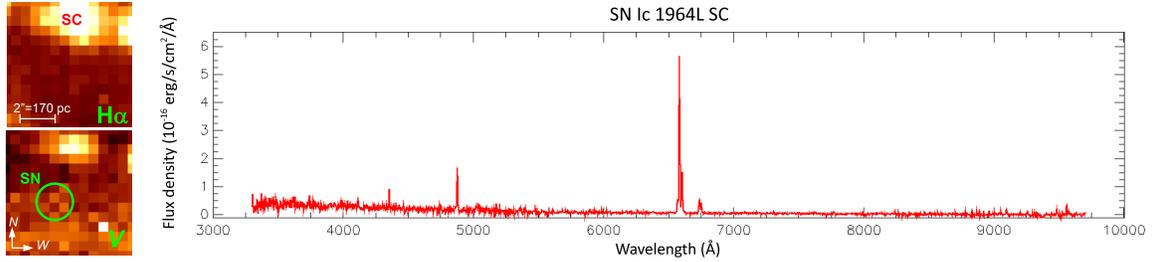}
\caption{Left panel: SN 1964L site reconstructed IFU FoV. SN position within 1 arcsec error radius is indicated by a circle. Approximate linear scale corresponding to 2 arcsec is also indicated; this scale is calculated from the host galaxy distance only thus does not take into account the projection effect and host galaxy inclination. "SC" indicates the host star cluster. Right panel: extracted spectrum of the star cluster.}
\label{sp64l}
\end{figure*}

\medskip
\begin{flushleft}
\textit{1. \objectname{SN 1964L} site} \\
\end{flushleft}
SN 1964L exploded in NGC 3938, a spiral galaxy hosting two other SNe (type II SNe 1961U and 2005ay -- see Paper II). This SN is poorly studied and no reference discussing the nature of the progenitor of this particular object was found in the literature. \citet{blaylock00} recovered the photographic spectra of the SN and classified it as type-Ic. We found that the cited position of SN 1964L is within 3 arcsec of a bright cluster in the host galaxy. Based on the measurements of the Palomar survey plates, the position was shown to be accurate to within 2" by \citet{porter93}. SNIFS data shows that this source exhibits a blue continuum with strong emission lines, typical of a young stellar population. Figure \ref{sp64l} shows the appearances of the explosion site on the IFU focal plane and the extracted spectrum of the host cluster. The measured metallicity is 1.09 solar. Measurement of H$\alpha$ equivalent width shows that the cluster is very young, with derived age of 3.3 Myr. The corresponding stellar initial mass of that age at solar metallicity is about 120~M$_\odot$. 

\medskip
\begin{flushleft}
\textit{2. \objectname{SN 1994I} site} \\
\end{flushleft}
SN 1994I is one of the best-studied type-Ic SNe due to its location in M51, a very well-observed nearby galaxy. It has often been referred as the 'standard' SN Ic although it also exhibited unusual behaviour \citep[e.g.][]{sauer06}. \citet{nomoto94} suggests that the progenitor of this SN is a 4.0~M$_\odot$ helium star in a binary system rather than a single WR star, which later evolved into a C+O star and eventually produced a SN Ic explosion. The main sequence initial mass of this star was estimated as 15~M$_\odot$. The position of this SN is accurate within 1", as estimated by \citet{vandyk92} for SNe of that era, and later confirmed by the late-time observations of \citet{clocchiatti08}.

\begin{figure*}
\plotone{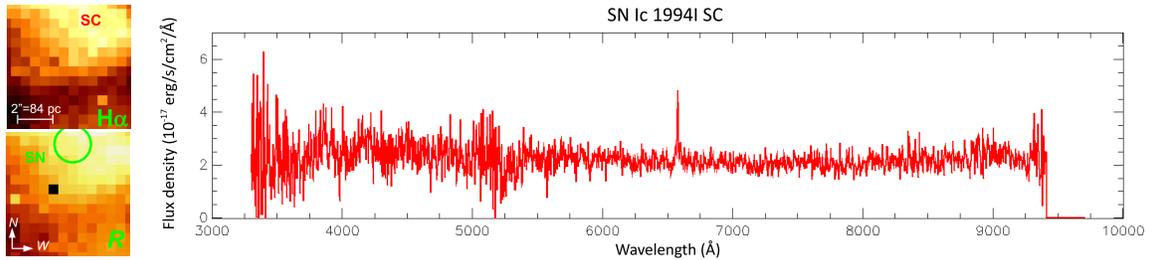}
\caption{IFU FoV and extracted host cluster spectrum for the SN 1994I site. The red end of the spectrum is truncated due to DAR. Figure annotations are the same as in Figure \ref{sp64l}.}
\label{sp94i}
\end{figure*}

Our SNIFS result shows that the explosion site is rather diffuse, the host cluster does not stand out prominently over the stellar background (Figure~\ref{sp94i}). The extracted spectrum shows faint continuum and a weak H$\alpha$ line. Metallicity was derived to be 0.83 solar from N2 -- this SN is the only SN Ic without an O3N2 determination in our sample since the lines H$\beta$ and [O\texttt{III}]$\lambda5007$ are not detected. The H$\alpha$ equivalent width corresponds to an age of 11.0 Myr. This age at solar metallicity corresponds to the lifetime of a 17.9~M$_\odot$ star, which is too low for a single WR progenitor but quite consistent with the binary model of \citet{nomoto94}. This is possibly one of the first strong pieces of evidence of a binary progenitor channel in SN Ic production.

\medskip
\begin{flushleft}
\textit{3. \objectname{SN 2000ew} site} \\
\end{flushleft}
SN 2000ew exploded in NGC 3810 and is the first SN Ic with detected emission of near-infrared carbon monoxide (CO) in the spectrum \citep{gerardy02}. \citet{vandyk03} searched the progenitor star in the pre-explosion images but found none within their detection limit. Comparing with HST images taken while the SN is still visible, the derived SN position is accurate to within 0.7". However, they discovered that luminous blue-yellow stars are present in the explosion site, within 1 arcsec of the SN position. The color-magnitude diagram suggests that the environment is very young, $\lesssim6$ Myr old. Similarly, \citet{maund05} also failed to detect the progenitor star in pre-explosion images and derived an age estimate of the environment within 6 arcsec from the SN of $\sim7$ Myr, assuming twice solar metallicity. \citet{maund05} estimated progenitor mass between 12 and 40~M$_\odot$ from the environment age.

\begin{figure*}
\plotone{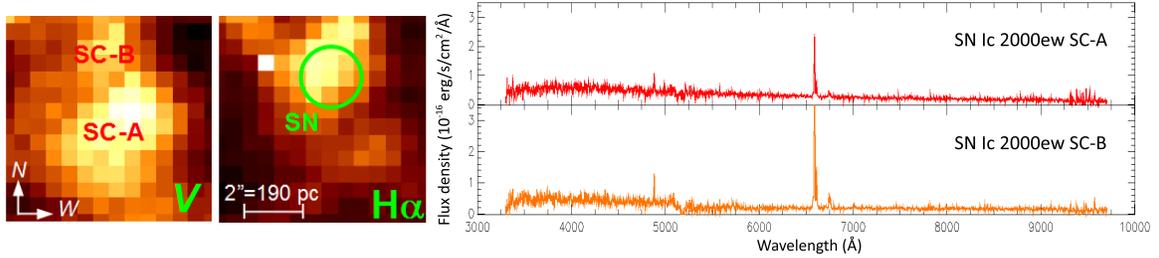}
\caption{IFU FoV and extracted cluster spectra for the SN 2000ew site. Figure annotations are the same as in Figure \ref{sp64l}.}
\label{sp00ew}
\end{figure*}

Our SNIFS result shows that the cluster coincident with SN position, indicated as SC-B in Figure \ref{sp00ew}, is indeed very young with age of 5.8 Myr. This is consistent with the estimate of \citet{vandyk03}. The brighter neighbouring cluster, SC-A, is slightly older, 6.3 Myr. Both clusters have similar metallicities of 1.15 and 1.05 solar, respectively. This on-site metallicity measurement is definitely more reliable compared to the twice-solar metallicity assumed in the previous studies. The corresponding main sequence initial mass for a 5.8 Myr stellar population at solar metallicity is 33.9~M$_\odot$, while it is 29.5~M$_\odot$ for the 6.3 Myr population.

\medskip
\begin{flushleft}
\textit{4. \objectname{SN 2004gt} site} \\
\end{flushleft}
\citet{galyam05}, and also \citet{maund05b} reported their efforts in recovering the pre-explosion progenitor of SN 2004gt in the Antennae Galaxy, NGC 4038. However, the detection results were negative and only upper luminosity limits could be derived for the progenitor star (positional uncertainty 5 mas). Both investigations could not rule out a massive Wolf-Rayet star progenitor, and also allow binary interaction scenario for producing the stripped-envelope progenitor star. The positional uncertainty of the SN in the HST images was estimated to be around 5 mas.

\begin{figure*}
\plotone{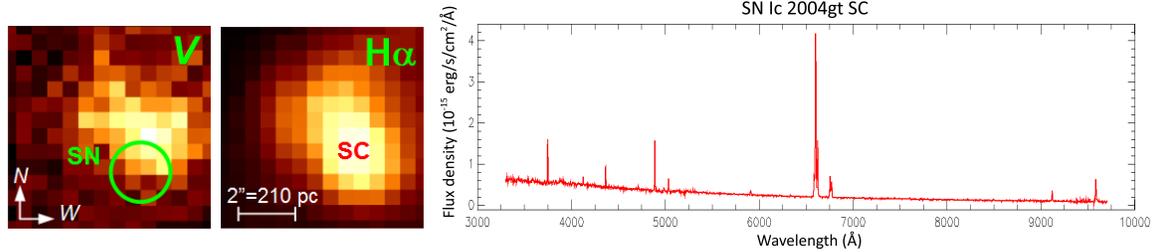}
\caption{IFU FoV and extracted host cluster spectrum for the SN 2004gt site. Figure annotations are the same as in Figure \ref{sp64l}.}
\label{sp04gt}
\end{figure*}

Our result with SNIFS shows that SN 2004gt exploded in a large star cluster complex (Figure \ref{sp04gt}), identified as Knot S by \citet{whitmore10}. We derived the metallicity of Knot S as 12+log(O/H) = 8.71, or 1.12 solar, with age of 5.78 Myr. This result is consistent with Knot S mean age estimate by \citet{whitmore10} of 5.75 Myr. The age of 5.78 Myr corresponds to the lifetime of a star with initial mass of 33.7~M$_\odot$ at solar metallicity. We also detect WR star feature in the spectra at 4650 \AA. A metallicity determination of the explosion site by \citet{modjaz11} gave 12+log(O/H) = $8.70^{+0.00}_{-0.01}$, using the PP04 O3N2 scale. Our O3N2 determination yields a higher value, 12+log(O/H) = 8.78. The difference may have come from the different regions being observed, due to the different spectroscopy techniques (slit vs. IFU). However, if the uncertainty in the PP04 O3N2 calibration of 0.14 dex is considered as the errors for both determinations, this difference is statistically insignificant, corresponding to only 0.4$\sigma$.

\medskip
\begin{flushleft}
\textit{5. \objectname{SN 2007gr} site} \\
\end{flushleft}
We used GMOS to observe the explosion site of SN 2007gr which exploded in the galaxy NGC 1058. \citet{hunter09} reported an extensive observation of the SN at optical and infrared wavelengths, covering more than 400 days. They showed that the photometric evolution of SN 2007gr is similar to SN 2002ap, an energetic Ic hypernova, but the spectra of the two SNe show marked differences. \citet{tanaka08} performed spectropolarimetric observation of the SN and suggested that a bipolar explosion viewed slightly off-axis may best represent the explosion model of SN 2007gr. A study using pre-explosion imaging data by \citet{crockett08} reported that the SN exploded very near ($\sim7$ pc) a point-like source in HST images, which they proposed as the compact host star cluster of the SN progenitor star. The positional uncertainty of the SN is about 0.02". Their determination of the compact cluster age using broadband photometry gives two solutions: 7 and 20--30 Myr assuming solar metallicity, corresponding to turnoff masses of 28 and 12--9~M$_\odot$, respectively. 

GMOS observation of the explosion site was seeing-limited at 0.5". Despite the excellent seeing condition, the compact clusters and bright individual stars comprising the OB association resolved in HST images could not be resolved and they appear collectively as bright knots in the IFU reconstructed images. We extracted the spectra of each of these five sources, designated SC-A through SC-E, including the central association where SN 2007gr exploded (SC-A). We derived the metallicity of SC-A as 1.12 solar, with age of 7.8 Myr, quite consistent with the 7~Myr solution of \citet{crockett08}. This age corresponds to the lifetime of a 24.4~M$_\odot$ star at solar metallicity. \citet{modjaz11} measured the metallicity at the explosion site to be 12 + log(O/H) = $8.64^{+0.07}_{-0.09}$ in PP04 O3N2 scale. Our O3N2 determination yields 12 + log(O/H) = 8.69, which is different with \citet{modjaz11}'s result only by 0.25$\sigma$ -- this is not significantly different.

Compared to other surrounding knots, SC-A appears to be older by about 1 Myr. All other knots SC-B through SC-E have ages between 6.4 and 6.8 Myr, which may suggest triggered star formation propagating inside-out from SC-A. If star formation in the outer clusters was triggered from SC-A, then the 1 Myr time difference and $\sim50$ pc projected distance would correspond to a projected propagation velocity of $\sim50$ km/s. The H$\alpha$ map of the region shows shell-like structure surrounding SC-A and coincident with the other four knots, leaving a depression in H$\alpha$ intensity at the position of SC-A -- further reinforcing this scenario. In any case, the ages of the other four knots corresponds very well with the lifetimes of massive stars, around 28--29~M$_\odot$. Interestingly, SC-E shows relatively lower, sub-solar metallicity (0.85~Z$_\odot$) compared to other knots present at the site, which are super-solar ($>$ 1.1 Z~$_\odot$).
\begin{figure*}
\plotone{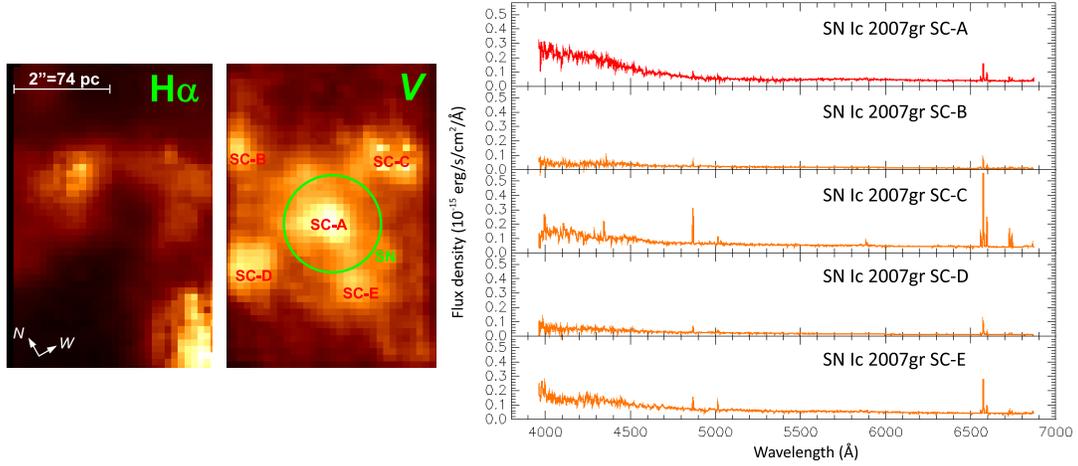}
\caption{IFU FoV and extracted cluster spectra for the SN 2007gr site. The continuum image of the GMOS FoV was made by collapsing the datacube in wavelength direction, approximately covering the $V$-band. Figure annotations are the same as in Figure \ref{sp64l}.}
\label{sp07gr}
\end{figure*}

\medskip
\begin{flushleft}
\textit{6. \objectname{SN 2009em} site} \\
\end{flushleft}
The Ic SN 2009em is not well studied, discovered by \citet{monard09} about 2 months after maximum light as confirmed by \citet{navasardyan09} and also by \citet{folatelli09}. The spectrum matches normal type Ic events. There is no other information on this object available in the literature. We assign positional uncertainty of the SN as 1", considering that it is a modern SN. The host galaxy NGC 157 is a grand-design Sc spiral, harbouring no other SN to date except 2009em.

With GMOS we observed the explosion site and detect two clusters. The more prominent cluster, SC-A, exhibits sub-solar metallicity of 0.76 Z$_\odot$, with age of 6.8 Myr. The neighbouring fainter cluster SC-B is barely detected in continuum images but stands out in H$\alpha$. It has nearly twice the metallicity of SC-A, 1.45 solar. The age of this cluster was derived as 6.3 Myr. The significant difference of the two clusters' metallicities illustrates the importance of measuring the metallicity of explosion sites with fine spatial resolution. The SN itself exploded at about the same distance from both clusters, but we assigned SC-A as the host cluster since it is the brighter one and thus more likely to host more progenitor candidates. The ages of the clusters correspond to the lifetimes of 27.7 and 29.5~M$_\odot$ stars, for SC-A and SC-B respectively.
\begin{figure*}
\plotone{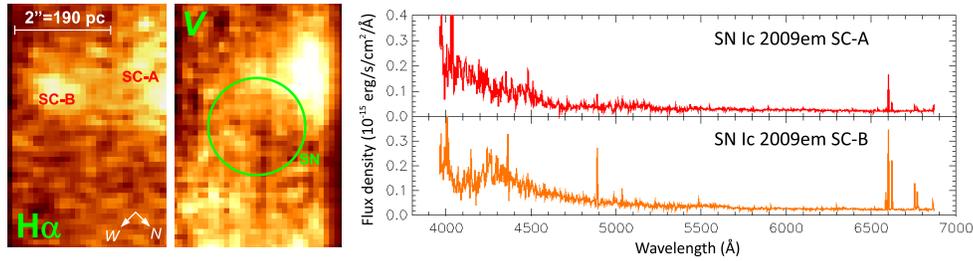}
\caption{IFU FoV and extracted cluster spectra for the SN 2009em site. Figure annotations are the same as in Figure \ref{sp64l}.}
\label{sp09em}
\end{figure*}

\begin{deluxetable*}{lcccccccrc}
\tabletypesize{\scriptsize}
\tablecaption{Results for SN Ic sites}
\tablewidth{0pt}
\tablehead{
\colhead{SN site} & \colhead{Object} & \colhead{Offset/"(pc)\tablenotemark{a}} & \colhead{(O3N2)} & \colhead{(N2)} & \colhead{12+log(O/H)} & \colhead{Z/Z$_\odot$} & \colhead{H$\alpha$EW/\AA} & \colhead{Age/Myr} & \colhead{M$_{}$/M$_\odot$}
}
\startdata
1964L & SC* & 2.6 (220) & 8.83  & 8.57 & 8.70 & $1.09^{+0.39}_{-0.27}$ & $939.10\pm84.52$  & $3.28^{+0.04}_{-0.04}$ & $120_{+1}^{-10}$ \\
1994I & SC* & 2.6 (110) & -- & 8.58 & 8.58 & $0.83^{+0.43}_{-0.28}$  & $13.34\pm1.87$ & $11.0^{+0.25}_{-0.75}$ & $17.9_{+0.8}^{-0.4}$ \\
2000ew & SC-A & 2.3 (220) & 8.77 & 8.58 & 8.68 & $1.05^{+0.23}_{-0.22}$ & $68.18\pm8.18$  & $6.31^{+0.04}_{-0.04}$  & $29.5_{+0.1}^{-0.2}$ \\
2000ew & SC-B* & 1.2 (110) & 8.85 & 8.59 & 8.72 & $1.15^{+0.40}_{-0.30}$ & $221.10\pm35.38$ & $5.75^{+0.09}_{-0.09}$ & $33.9^{-0.8}_{+1.0}$ \\
2004gt & SC* & 1.3 (140) & 8.78 & 8.64 & 8.71 & $1.12^{+0.20}_{-0.17}$ & $209.60\pm10.48$ & $5.78^{+0.03}_{-0.03}$ & $33.7^{-0.3}_{+0.2}$ \\
2007gr & SC-A* & 0.0 (0) & 8.69 & 8.72 & 8.71 & $1.12^{+0.03}_{-0.05}$ & $15.57\pm0.96$ & $7.84^{+0.75}_{-0.25}$  & $24.4_{+0.8}^{-2.5}$ \\
2007gr & SC-B & 1.4 (52) & 8.96 & 8.63 & 8.80 & $1.35^{+0.64}_{-0.42}$ & $33.03\pm3.30$ & $6.66^{+0.06}_{-0.05}$ & $28.3_{+0.1}^{-0.2}$ \\
2007gr & SC-C & 1.1 (41) & 8.82 & 8.68 & 8.75 & $1.20^{+0.25}_{-0.20}$ & $59.62\pm2.47$ & $6.36^{+0.01}_{-0.02}$ & $29.3_{+0.1}^{-0.1}$ \\
2007gr & SC-D & 1.6 (59) & 8.99 & 8.58 & 8.79 & $1.35^{+0.79}_{-0.52}$ & $50.26\pm5.03$ & $6.42^{+0.06}_{-0.04}$ & $29.1_{+0.1}^{-0.2}$ \\
2007gr & SC-E & 1.0 (37) & 8.65 & 8.54 & 8.60 & $0.85^{-0.14}_{-0.09}$ & $27.95\pm1.39$ & $6.78^{+0.08}_{-0.05}$ & $27.9_{+0.2}^{-0.3}$ \\
2009em & SC-A* & 1.1 (100) & 8.57 & 8.51 & 8.54 & $0.76^{+0.05}_{-0.05}$ & $26.88\pm2.69$ & $6.84^{+0.30}_{-0.09}$ & $27.7_{+0.4}^{-1.0}$ \\
2009em & SC-B & 1.0 (94) & 8.85 & 8.78 & 8.82 & $1.45^{+0.10}_{-0.13}$ & $68.76\pm5.09$ & $6.31^{+0.02}_{-0.03}$ & $29.5_{+0.1}^{-0.1}$ \\
\enddata
\tablenotetext{a}{Offset between SN and approximate cluster center.}
\tablenotetext{*}{SN parent cluster.}
\label{tab1c}
\end{deluxetable*}

\medskip
\subsection{SNe Ib sites}

\medskip
\begin{flushleft}
\textit{1. \objectname{SN 1983N} site} \\
\end{flushleft}
SN 1983N exploded in a nearby large spiral galaxy, M83 (NGC 5236). This SN was observed extensively \citep[e.g.][]{gaskell86, weiler86, clocchiatti96} and has been suggested to be the prototypical type Ib SNe along with SN 1984L \citep{porter87}. \citet{sramek84} reported the first radio detection of a type I SN of this SN 1983N. \citet{clocchiatti96} reported that their astrometry of the SN position is accurate to within $\sim0.6$".

\begin{figure*}
\plotone{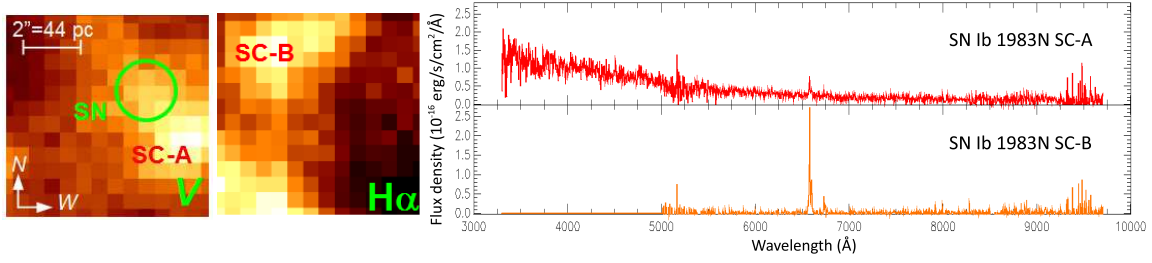}
\caption{IFU FoV and extracted cluster spectra for the SN 1983N site. Figure annotations are the same as in Figure \ref{sp64l}. The blue part of SC-B spectrum could not be extracted since it fell outside the FoV due to differential atmospheric refraction.}
\label{sp83n}
\end{figure*}

Using SNIFS we discovered that the explosion site of the SN is quite complex, with three detected objects within the IFU FoV. In the continuum a prominent cluster was detected, but this cluster is overshadowed in H$\alpha$ by two bright nearby H\texttt{II} regions (see Figure \ref{sp83n}). We extracted the spectra of SC-A cluster and the H\texttt{II} region north-east of it (SC-B), but failed to do so for the H\texttt{II} region south-east of the cluster since it is situated at the edge of the FoV. It is likely that this third cluster is similar in age with SC-B considering its detectability in H$\alpha$ and nondetection in continuum images. The cluster SC-A was found to be 7.2 Myr old with 1.02 times solar metallicity while the H\texttt{II} region SC-B has age of 2.6 Myr with 0.89 solar metallicity. The age of SC-A corresponds to the lifetime of a 26.4~M$_\odot$ star while the SC-B age is considered too young to even produce a supernova.

\medskip
\begin{flushleft}
\textit{2. \objectname{SN 1984L} site} \\
\end{flushleft}
SN 1984L exploded in NGC 991, a low surface brightness spiral galaxy. \citet{schlegel89} presented late-time photometry and spectroscopy of the SN, and suggested that it originated from a roughly 20~M$_\odot$ star. Later, \citet{swartz91} derived an ejecta mass of more than 10~M$_\odot$, which they inferred to be the result of an explosion of a $\gtrsim30$~M$_\odot$ star. \citet{baron93} put forward the problem in reconciling the early-time light curve with the spectra of the SN, for which they proposed a solution by suggesting the scenario in which explosion of a very massive star ($\sim50$~M$_\odot$ of helium with 2 $\times10^{52}$ erg kinetic energy) produced a massive black hole. 
The coordinates given by \citet{schlegel89} differ by about 0.3" compared to those of \citet{weiler86}. This value is used as the positional uncertainty of SN 1984L.

Our GMOS observation of the explosion site shows that the host star cluster of SN 1984L does not exhibit strong emission lines in its spectrum, indicating intermediate age. We measured the metallicity to be 0.45 solar, the lowest in our sample. H$\alpha$ emission equivalent width is small, giving an age estimate of 18.0 Myr, which corresponds to the lifetime of a 13.5~M$_\odot$ star. This result is inconsistent with previous studies suggesting a high-mass ($\gtrsim20$~M$_\odot$) progenitor, but intriguing since the derived mass of 13.5~M$_\odot$ lies well below the currently accepted WR star mass limit and within the range of SN II-P progenitors. With such low metallicity, it is even more difficult to remove the hydrogen envelope of a single star via stellar wind mechanism, and as a result the minimum mass of a star to become a WR star increases. Thus, this is a strong indication that the SN 1984L progenitor may have lost its envelope in a close binary system.
\begin{figure*}
\plotone{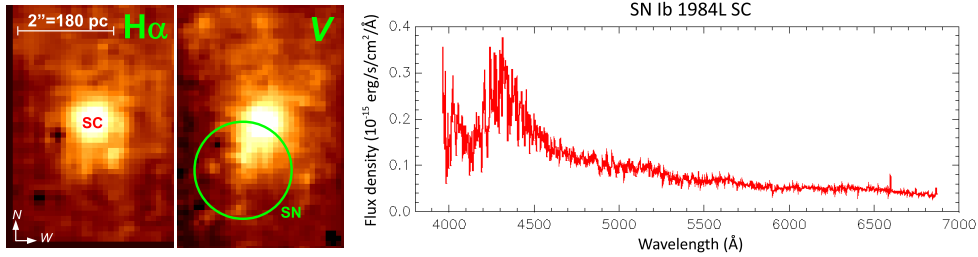}
\caption{IFU FoV and extracted host cluster spectrum for the SN 1984L site. Figure annotations are the same as in Figure \ref{sp64l}.}
\label{sp84L}
\end{figure*}

\medskip
\begin{flushleft}
\textit{3. \objectname{SN 1999ec} site} \\
\end{flushleft}
The Ib SN 1999ec exploded in the interacting spiral galaxy, NGC 2207. Currently there have been four recorded SNe that have occurred in this galaxy (SNe 1975A, 1999ec, 2003H, 2013ai). \citet{vandyk03} searched for the progenitor star in pre-explosion images but found no convincing candidate. Their HST images show that the SN occurred in an environment of blue, apparently young stars and clusters. The SN position uncertainty was $\pm0.2$".

\begin{figure*}
\plotone{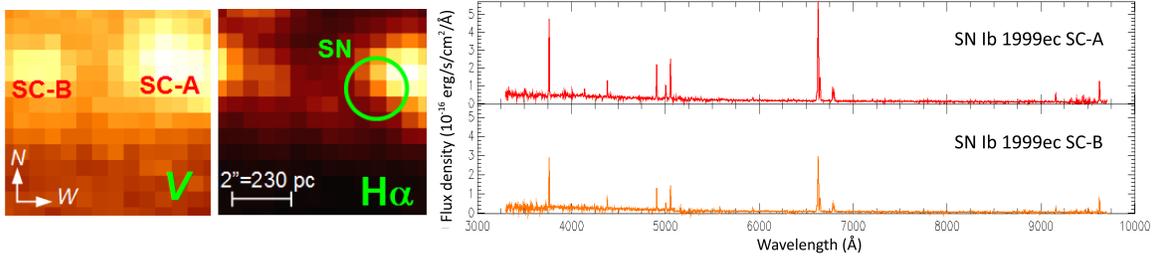}
\caption{IFU FoV and extracted cluster spectra for the SN 1999ec site. Figure annotations are the same as in Figure \ref{sp64l}.}
\label{sp99ec}
\end{figure*}

With SNIFS we detect two large clusters in the SN vicinity (Figure \ref{sp99ec}). Both clusters have similar spectra and are of similar metallicity, with SC-A being 0.71 solar value and SC-B of 0.74 solar. Also in derived age, both show similarity, with SC-A being 5.3 Myr and SC-B being 5.1 Myr. These age estimates at solar metallicity correspond to the lifetimes of 38.0 and 41.6~M$_\odot$ stars, respectively. The derived initial mass is high, even compared to the majority of SN Ic progenitors.

\medskip
\begin{flushleft}
\textit{4. \objectname{SN 2008bo} site} \\
\end{flushleft}
SN 2008bo exploded in a spiral host, NGC 6643. This SN has been detected in X-ray \citep{immler08,immler08b} and radio \citep{stockdale08} wavelengths. The Asiago database assigned the type Ib for this SN; \citet{navasardyan08} reported that the spectrum shows the characteristics of a stripped-envelope event but with deep H$\alpha$ absorption. This resembles the spectrum of SN 2008ax, a type IIb SN, which shows spectral evolution from type II to Ib. There is no other useful discussion about the physical nature of the object in the literature. \citet{stockdale08} reported that the radio position of this SN agrees well ($\sim0.2$") with the optical position. Thus, we assign 0.2" as the positional uncertainty of this SN.

\begin{figure*}
\plotone{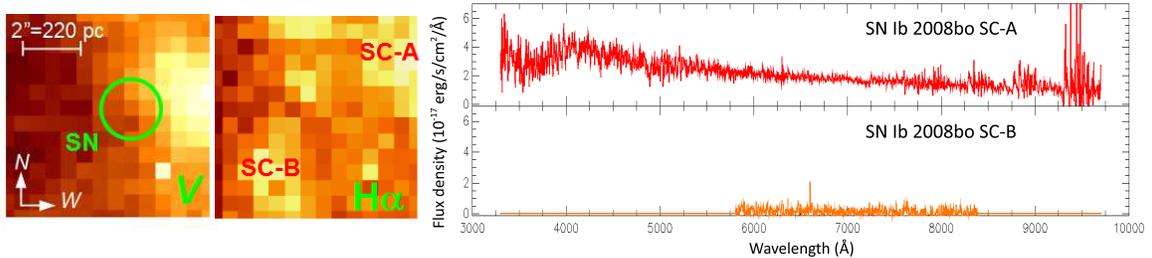}
\caption{IFU FoV and extracted cluster spectra for the SN 2008bo site. Figure annotations are the same as in Figure \ref{sp64l}. Only a part of SC-B spectrum could be recovered due to DAR effects.}
\label{sp08bo}
\end{figure*}

We observed the explosion site using SNIFS and extracted the spectra of two clusters apparent at the site (Figure \ref{sp08bo}). A prominent cluster, SC-A, appears in all wavelengths while in H$\alpha$ we detect a faint secondary cluster, SC-B, and some other fainter structures in the FoV. We performed additional imaging observation of the SN 2008bo site using OPTIC camera at UH88 \citep{howell03} in 2011. However, the surrounding fainter structures are not visible in our broadband images. From SNIFS data SC-A was found to be 13.5 Myr old with 1.20 solar metallicity. We only managed to extract a part of the spectrum of SC-B but nevertheless recovered the H$\alpha$ region and derived an age via its equivalent width, which is 6.4 Myr. The metallicity of SC-B is derived as 0.85 solar. The 13.5 Myr age corresponds to a star with initial mass of 14.9 M$\odot$, while 6.4 Myr corresponds to 29.3~M$_\odot$. The 14.9~M$_\odot$ derived mass is too small for a star to produce a WR star that may later explode as a SN Ib, thus it is likely that SN 2008bo also may have been produced by a progenitor in binary system.

\medskip
\begin{flushleft}
\textit{5. \objectname{SN 2009jf} site} \\
\end{flushleft}
SN 2009jf exploded in one of the two prominent spiral arms of the barred spiral galaxy NGC 7479. It was an energetic slow-evolving SN, indicating massive ejecta. \citet{sahu11} presented 9-months of optical photometry and spectroscopy monitoring of this SN and suggested the main sequence mass of the progenitor to be $\gtrsim$ 20--25~M$_\odot$ estimated from the ejecta mass. Further, \citet{valenti11} also reported their 1-year monitoring of SN 2009jf and similarly concluded that the progenitor mass range is within 25--30~M$_\odot$. Their positional uncertainty of the SN is in the order of 0.06".

\begin{figure*}[ht!]
\plotone{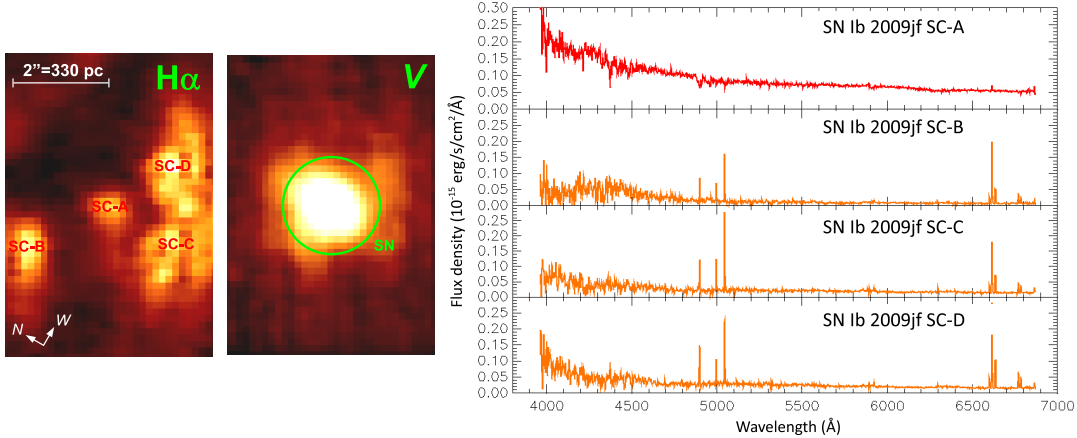}
\caption{IFU FoV and extracted cluster spectra for the SN 2009jf site. Figure annotations are the same as in Figure \ref{sp64l}.}
\label{sp09jf}
\end{figure*}

We observed the explosion site of SN 2009jf using GMOS, two years after the discovery. The continuum IFU reconstructed image shows only one prominent source in the field (Figure \ref{sp09jf}), which are the blended sources A+B in \citet{valenti11}. We assigned the designation SC-A star cluster for this blended source. Making use of the capability of IFU we reconstructed H$\alpha$ image of the field, and found that clumpy shell-like structure of ionized gas is present around SC-A. We assigned identification SC-B, SC-C, and SC-D for these clumps. We found that the metallicity of this complex is below solar, with SC-A being 0.78 solar value and SC-B, C, D within 0.63--0.81 solar. The N2 metallicity of SC-A in 12+log(O/H) equals 8.55 dex, agrees exactly with the explosion site metallicity estimate of \citet{sanders12}, using slit spectroscopy of 0.7" slit width. However, they do not mention about the aperture size for spectrum extraction nor the seeing size during the observation. \citet{sanders12} determined the age of the explosion site from H$\beta$ equivalent width to be 9.0 Myr. The age of SC-A from our determination is quite old, 18.2 Myr which corresponds to the lifetime of a 12.4~M$_\odot$ star. The surrounding clusters are significantly younger, between 6.0 and 6.4 Myr. These morphological characteristics are similar with what we found at the SN 2007gr explosion site, where the central host cluster in the complex is older than the surroundings thus suggesting inside-out star formation activity. This formation scenario has been observed elsewhere \citep[e.g.][]{adamo12}, thus suggesting that triggered star formation may not be uncommon in star formation complexes. Assuming the outer clusters formation was triggered from SC-A, the projected outward velocity of star formation propagation in this complex is $\sim300$ pc per 12 Myr, or around 25 km/s. This is about half the projected velocity we inferred for the SN 2007gr site.  

If the progenitor of SN 2009jf indeed originated in SC-A, this would be a third example after SN 1984L and SN 2008bo of low-mass SN progenitor produced by binary channel. In contrast, the surrounding environment age is consistent with the lifetime of stars with mass around 30~M$_\odot$. We note that the progenitor mass inferred from SN properties is more consistent with the outer environment age rather than the stellar association closest with the SN position. \citet{valenti11} found that the color of their source A is consistent with a stellar population containing stars with maximum stellar mass of 8--25~M$_\odot$, while source B is redder thus presumably older. This is not inconsistent with our findings which yield progenitor mass of about 12~M$_\odot$. However, it is possible that the star formation history in the immediate vicinity of SN 2009jf is not instantaneous, and the progenitor star was born from a younger burst compared to the dominant older surrounding population.

\begin{deluxetable*}{lcccccccrc}
\tabletypesize{\scriptsize}
\tablecaption{Results for SN Ib sites}
\tablewidth{0pt}
\tablehead{
\colhead{SN site} & \colhead{Object} & \colhead{Offset/"(pc)\tablenotemark{a}} & \colhead{(O3N2)} & \colhead{(N2)} & \colhead{12+log(O/H)} & \colhead{Z/Z$_\odot$} & \colhead{H$\alpha$EW/\AA} & \colhead{Age/Myr} & \colhead{M$_{}$/M$_\odot$}
}
\startdata
1983N & SC-A* & 2.6 (57) & -- & 8.67 & 8.67 &  $1.02^{+0.53}_{-0.34}$ & $23.32\pm4.19$ & $7.22^{+0.27}_{-0.42}$ & $26.4_{+1.4}^{-0.8}$ \\
1983N & SC-B & 3.0 (66) &  -- & 8.61 & 8.61 & $0.89^{+0.46}_{-0.30}$ & $1693\pm728$ & $2.62^{+0.65}_{-1.26}$ & -- \\
1984L & SC* & 1.6 (150) & -- & 8.31 & 8.31 & $0.45^{+0.13}_{-0.16}$ & $2.93\pm0.49$ & $18.0^{+2.18}_{-2.21}$ & $13.5_{+1.1}^{-0.9}$ \\
1999ec & SC-A* & 0.8 (90) & 8.49 & 8.53 & 8.51 & $0.71^{+0.30}_{-0.30}$ & $384.50\pm46.14$ & $5.34^{+0.15}_{-0.35}$ & $38.0_{+5.0}^{-1.5}$ \\
1999ec & SC-B & 3.3 (370) & 8.51 & 8.55 & 8.53 & $0.74^{+0.04}_{-0.02}$ & $424.10\pm72.09$ & $5.06^{+0.40}_{-0.71}$ & $41.6_{+14.2}^{-4.8}$ \\
2008bo & SC-A* & 3.1 (330) & -- & 8.74 & 8.74 & $1.20^{+0.62}_{-0.41}$ & $3.13\pm0.53$ & $13.5^{+1.03}_{-0.40}$ & $14.9_{+0.4}^{-0.5}$ \\
2008bo & SC-B & 3.5 (380) & -- & 8.59 & 8.59 & $0.85^{+0.44}_{-0.29}$ & $56.37\pm48.48$ & $6.37^{+5.92}_{-0.24}$ & $29.3_{+0.9}^{-13.0}$ \\
2009jf & SC-A* & 0.0 (0) & -- & 8.55 & 8.55 & $0.78^{+0.39}_{-0.27}$ & $1.38\pm0.12$ & $18.2^{+0.77}_{-0.52}$ & $12.4_{+0.3}^{-0.5}$ \\
2009jf & SC-B & 1.4 (230) & 8.42 & 8.53 & 8.48 & $0.66^{+0.08}_{-0.08}$ & $140.6\pm23.90$ & $6.02^{+0.16}_{-0.16}$ & $34.7_{+1.2}^{-0.9}$ \\
2009jf & SC-C & 1.0 (160) & 8.39 & 8.52 & 8.46 & $0.63^{+0.09}_{-0.09}$ & $88.71\pm9.76$ & $6.36^{+0.13}_{-0.08}$ & $32.2_{+0.6}^{-1.0}$ \\
2009jf & SC-D & 1.2 (200) & 8.50 & 8.63 & 8.57 & $0.81^{+0.12}_{-0.12}$ & $80.36\pm9.64$ & $6.25^{+0.04}_{-0.05}$ & $29.7_{+0.1}^{-0.2}$\\
\enddata
\tablenotetext{a}{Offset between SN and approximate cluster center.}
\tablenotetext{*}{SN parent cluster.}
\label{tab1b}
\end{deluxetable*}


\bigskip
\section{Discussions}
\label{discu}

It is very important to determine the metallicity and age of the explosion sites of SNe Ib/c to characterize their progenitor stars. Metallicity-driven wind is one widely accepted scenario to remove the star's hydrogen envelope and produce a stripped SN. Our results show that even at the SN explosion site metallicity variations from place to place may be present, thus emphasizing the importance of measuring local metallicity directly with the highest spatial resolution possible and discouraging metallicity determinations using proxies. The results are presented in Tables \ref{tab1b} and \ref{tab1c}.

Averaging metallicities of all the 12 objects present at each explosion sites, we found mean metallicity for the sample of SN Ic sites as $1.11\pm0.22$ (RMS; standard error of the mean/SEM = $\sigma/\sqrt{N}$ = $\pm0.06$) Z$_\odot$, while this value is reduced to $1.01\pm0.17$ (SEM = $\pm0.08$) Z$_\odot$ if only the SN parent clusters are considered. For the case of SN Ib sites the mean metallicity value is $0.79\pm0.20$ (SEM = $\pm0.06$) Z$_\odot$, and $0.83\pm0.29$ (SEM = $\pm0.13$) Z$_\odot$ for SN Ib host clusters only. Notwithstanding the somewhat small number of samples (6 SN Ic sites, 12 clusters; 5 SN Ib sites, 11 clusters), this immediately suggests that SNe Ic are produced in higher metallicity environments than SNe Ib. The statistical significance of the difference between SN Ic and Ib host clusters metallicities determined by the $t$-test is 1.2$\sigma$, while it is 3.7$\sigma$ if we consider the site metallicities. 

\citet{modjaz11} argued that SNe Ic and Ib explode in different metallicity environments with a difference of $0.2$ dex in 12+log(O/H), with total number of sample of $N = 35$. Their Kolmogorov-Smirnov (KS) test p-value of the null hypothesis of both SN types came from the same population yields $1$\%. Converting back our solar-unit metallicity values of the host clusters into 12+log(O/H) scale, we found $0.08$ dex difference between SN Ic and Ib explosion sites. The KS test p-value for our two samples is $59$\%, which could not rule out the possibility that both samples are from the same population -- however this might not be very meaningful considering the small number statistics. For comparison, other studies by \citet[][$N = 20$]{leloudas11} and \citet[][$N = 33$]{sanders12} also suggest higher metallicity for Ic (albeit not statistically significant), while \citet{anderson10} found no difference between Ib and Ic metallicities ($N = 24$). The sample of \citet{sanders12} was drawn from untargeted SN surveys, thus is likely to be less biased towards massive, high-metallicity host galaxies compared to a sample drawn from targeted SN surveys. Our result ($N = 11$) of $1.2\sigma$ significance suggests that on average SNe Ic are the more metal rich compared to SNe Ib, but not statistically significant. Therefore, it is more closer to the conclusion of \citet{leloudas11} who also found suggestion that SNe Ic are more metal rich compared to SNe Ib, similarly by $0.08$ dex.


We note one interesting fact that for SN Ic sites the metallicity derived from O3N2 index is almost always (10 out of 11) higher than from N2 index ($3.7\sigma$ significance), while it is the opposite for SN Ib sites ($2.7\sigma$ significance). Consistently for both SN Ic and Ib sites the measured metallicity is always higher for SN Ic sites in either O3N2 or N2 scales. The high O3N2 metallicity may be indicative of N/O enhancement \citep{perezmontero09}. This behavior has been observed in galaxies with high O3N2 metallicity as reported by \citet{berg11}, and interpreted as galaxies evolving past Wolf-Rayet galaxy phase as WR stars may elevate the observed N/O ratio.

The age of the SN explosion site provides clue to the likely mass of the progenitor star. We averaged the age of all detected clusters in our SN sites and found that SN Ic sites have younger mean age than SN Ib sites, $6.6\pm1.7$ (SEM = $\pm0.5$) Myr compared to $8.6\pm5.3$ (SEM = $\pm1.6$) Myr. This trend is also present when comparing the ages of the host clusters, with SN Ic hosts averaged $6.7\pm2.6$ (SEM = $\pm1.1$) Myr while for SN Ib hosts the age is significantly higher, $12.4\pm5.9$ (SEM = $\pm2.6$) Myr. This result shows that it is likely that SNe Ic resulted from higher-mass progenitors than SNe Ib (2.0$\sigma$ significance). Considering the results for individual objects, we also suggest that some SN Ib/c progenitors may have originated from close binary systems due to the derived old age of the host cluster. SN 1994I is one example for SN Ic with derived old parent cluster age, or low mass progenitor. The derived age of 10.9 Myr corresponds to the lifetime of a 17.9~M$_\odot$ single star, which is not massive enough to evolve into a stripped-envelope WR star thought necessary to produce SN Ic. Therefore, one likely scenario is that this SN was the explosion of a star in a binary system, whose hydrogen envelope has been removed by binary interactions. For SNe Ib in our sample we found that this scenario may be applicable to SNe 1984L, 2008bo, and 2009jf. It is also interesting to note that the binary scenario is significantly more prevalent in our SNe Ib samples (3 out of 5 SNe) compared to SNe Ic (1 out of 6 SNe), but further investigation with more samples is necessary to confirm this indication firmly. 

It is interesting to note that type IIb SNe, whose spectral evolution showed transition from type II to type Ib, already show evidence that points to a star in a binary system as the progenitor. \citet{aldering94} showed that the SED of SN 1993J progenitor could not be fit with a single star SED; further, \citet{maund04} reported the unambiguous detection of the companion star with post-explosion observation. Indication of binarity was also found in recent IIb events: SN 2008ax \citep{crockett08b} and SN 2011dh \citep{bersten12}. Based on the similarities, it is possible that both type IIb and Ib SNe have similar binary progenitors.

\begin{figure*}
\epsscale{1.1}
\plotone{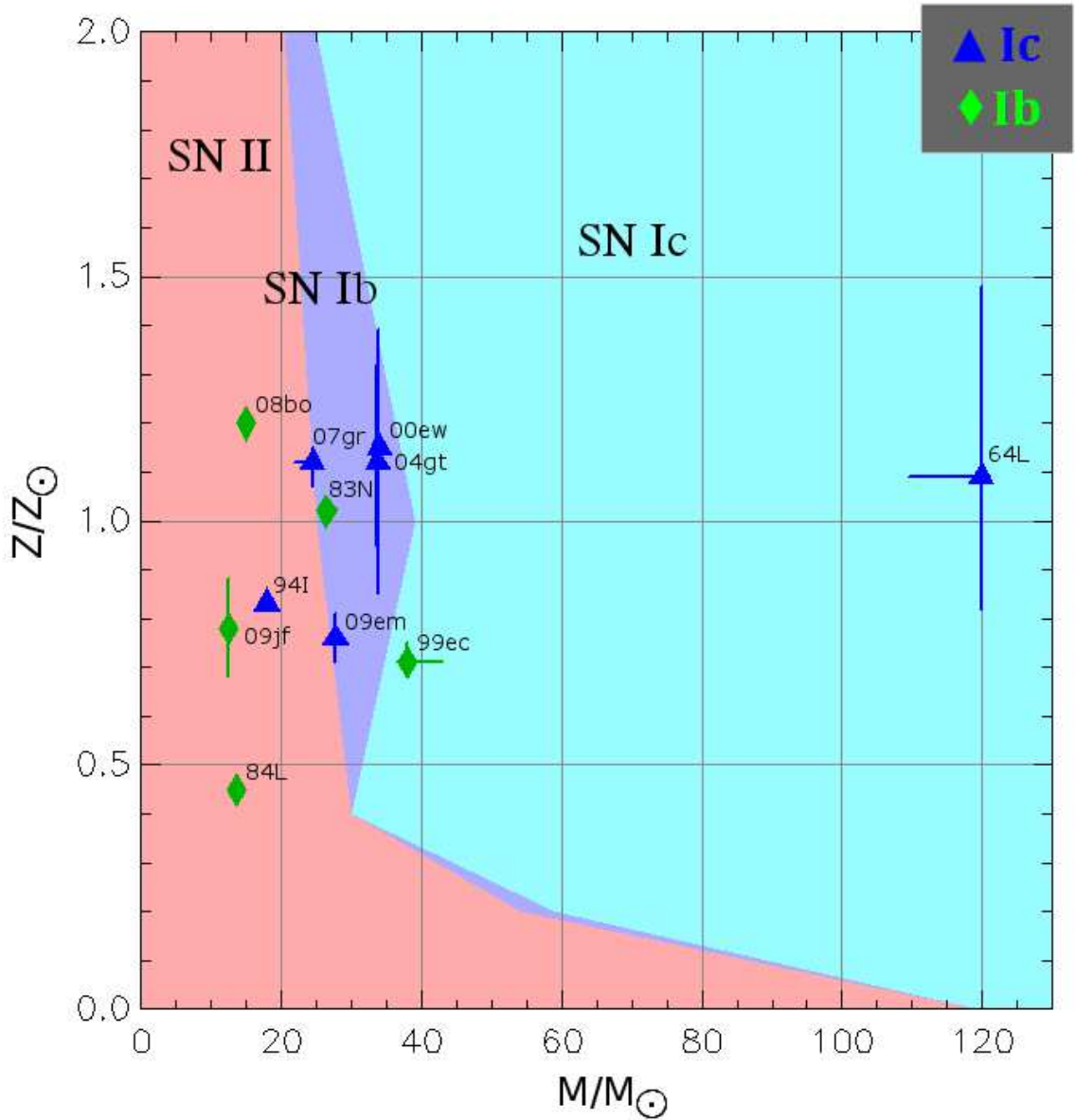}
\caption{SN Ib/c progenitors overplotted on mass-metallicity space; Ic progenitors are indicated with purple triangles and Ib progenitors are indicated with green diamonds. Theoretical predictions of progenitors of each SN type by \citet{georgy09} are drawn as colored regions.}
\label{mzdiag}
\end{figure*}

With the information on SN host cluster age and metallicity, we derive the maximum mass of the stars still present in the cluster. As the most massive stars dies first, any of the remaining stars would not exceed the mass of the SN progenitor exploded just recently. The age of the cluster is equal to the lifetime of the SN progenitor, and since the lifetime of a single star is mainly governed by its initial mass, the lifetime of the SN progenitor could be used to estimate its initial mass. With this method we determined the main-sequence mass of the SN progenitors in our sample by comparing to Padova stellar evolution models, adopting the metallicity of the parent cluster. We used the models by \citet{bressan93} for solar metallicity and \citet{fagotto94} for 0.4 solar metallicity. The resulting mass along with metallicity information for each SN progenitor is plotted on mass-metallicity space in Figure \ref{mzdiag}. Theoretical predictions of SN progenitors by \citet{georgy09} are overlaid to make the comparison readily available. It is immediately apparent that the theoretical models and observational data points do not agree very well -- only one occurrence of observational data falls into the right theoretical prediction for each Ib/c type (SNe 1983N and 1964L). In spite of this, we argue that the relative comparison of the mass of progenitor star of different SN types (i.e. Ib versus Ic) should be robust regardless of the stellar evolution model used. We also compare our results of progenitor mass and local environment age with other observational results using different methods, summarized in Table \ref{tabcomp}. In general our results do not contradict the findings of other studies, except probably for SN 1984L. This provides a strong confirmation of the validity of our method. However, we note that there is always probability that we are suffering from projection effects which may cause apparent association between the SN and an otherwise unrelated stellar population, resulting in a chance superposition. This issue is further addressed in Paper II, resulting in a 50\% chance superposition estimate. Nevertheless our method offers the current best determination of SN progenitor age and metallicity from the SN local environment. Median distance of objects in this study is $\sim$15 Mpc (1" corresponds to 70 pc), or redshift 0.0035 (for $H_0$ = 70 km s$^{-1}$ Mpc$^{-1}$), the smallest compared to other similar studies. For example the median redshift (distance) of objects in \citet{anderson10} is 0.005 (21 Mpc), \citet{modjaz11} 0.017 (72 Mpc), \citet{leloudas11} 0.022 (94 Mpc), \citet{sanders12} 0.036 (154 Mpc) \citep[Table 8 of ][]{sanders12}.

\begin{deluxetable*}{lcccccc}
\tabletypesize{\scriptsize}
\tablecaption{Mass and age of SN Ib/c progenitors from this work, compared with other determinations}
\tablewidth{0pt}
\tablehead{
\colhead{SN} & \colhead{Mass (this w.)}& \colhead{Age (this w.)} & \colhead{Mass (ref.)} & \colhead{Age (ref.)} & \colhead{Method (ref.)} & \colhead{Reference} 
}
\startdata
Ic 1964L & 120~M$_\odot$ & 3.28 Myr & -- & -- & -- & -- \\
Ic 1994I & 17.9~M$_\odot$ & -- & 15~M$_\odot$ & -- & SN properties & \citet{nomoto94} \\
Ic 2000ew & -- & 5.75 Myr & -- & $\lesssim6$ Myr & environment CMD & \citet{vandyk03} \\
 & 33.9~M$_\odot$ & -- & 12--40~M$_\odot$ & -- & nondetection & \citet{maund05} \\
Ic 2004gt & -- & 5.78 Myr & -- & 5.75 Myr & SC color & \citet{whitmore10} \\
 & 33.7~M$_\odot$ & -- & $\gtrsim40$~M$_\odot$, or & -- & nondetection & \citet{maund05b}, \\
  & & & low-mass binary & & & \citet{galyam05} \\
Ic 2007gr & 24.4 M$_\odot$ & 7.84 Myr & 28 or 12-9 M$_\odot$ & 7 or 20--30 Myr & SC color & \citet{crockett08} \\
Ic 2009em & 27.7~M$_\odot$ & 6.84 Myr & -- & -- & -- & -- \\
Ib 1983N & 26.4~M$_\odot$ & 7.22 Myr & --& -- & -- & --  \\
Ib 1984L & 13.5~M$_\odot$ & 18.0 Myr & 20~M$_\odot$ & -- & SN properties & \citet{schlegel89} \\
 & & & $\gtrsim30$~M$_\odot$ & -- & SN properties & \citet{swartz91} \\
 & & & $\gtrsim50$~M$_\odot$ & -- & SN properties & \citet{baron93} \\
Ib 1999ec & 38.0~M$_\odot$ & 5.34 Myr & $M_V\gtrsim-8.7$ & --&  nondetection & \citet{vandyk03} \\
Ib 2008bo & 14.9~M$_\odot$ & 13.5 Myr & -- & -- & -- & -- \\
Ib 2009jf & -- & 6--18 Myr & -- & 9.0 Myr & site H$\beta$EW & \citet{sanders12} \\
 & 12.4--34.7~M$_\odot$ & -- & $\gtrsim$20--25~M$_\odot$ & -- &SN properties & \citet{sahu11} \\
 & 12.4--34.7~M$_\odot$ & -- & 25--30~M$_\odot$ & -- &  SN properties & \citet{valenti11} \\
 & 12.4--34.7~M$_\odot$ & -- & $\lesssim$8--25~M$_\odot$ & -- &   SC color & \citet{valenti11} \\
\enddata
\tablecomments{CMD: color-magnitude diagram; SC: star cluster}
\label{tabcomp}
\end{deluxetable*}

In this work we suggest that SNe Ic originate from more massive progenitors than SNe Ib. The progenitor mass of SN Ic averages $42.9\pm38.2$~M$_\odot$, while for SN Ib the average progenitor mass is $21.0\pm11.0$~M$_\odot$. The high O3N2 metallicity of SN Ic sites supports this conclusion that SN Ic progenitors have mass within the WR star range. We note that the large deviation in SN Ic average progenitor mass is caused by the very high determined mass of SN 1964L, $\sim120$~M$_\odot$. If we omit this one SN the average progenitor mass for SN Ic is reduced to $27.5\pm6.7$~M$_\odot$ -- still higher than SN Ib progenitor average and consistent with WR star mass. The statistical significance of the Ib and Ic progenitor mass difference is $1.3\sigma$, and the inclusion of SN 1964L will slightly reduce it into $1.1\sigma$. Our results also show some examples where the progenitor may have originated from close binary systems, considering the old age of the parent stellar population. This is consistent with the findings of \citet{leloudas11}, which suggests that both single or binary progenitor scenarios are at work for SN Ib/c production. Recent finding by \citet{sana12} shows that binary interaction may affect 70\% or more of the massive star population. 

Considering the two channels of high and low mass progenitors possible for SN Ib/c, it is interesting to compare it with the derived initial masses of the compact remnants in the Galaxy. Magnetars are believed to be formed by rapidly rotating massive stellar cores, and this mechanism would not work if the star enters the red supergiant phase where magnetic breaking between the stellar core and envelope may spin down the core rotation. Therefore, a very high mass star ($\gtrsim40$~M$_\odot$) that can lose a significant portion of its hydrogen envelope prior to SN is required as the progenitor of magnetars. Several studies have confirmed this by studying the host star clusters of the magnetars \citep[e.g.][]{muno06}, but \citet{davies09} showed that the magnetar SGR 1900+14 may have had a low mass progenitor of around 17~M$_\odot$. Therefore, the SN that produced those magnetars should also came from both massive and lower-mass stars, and the SN needs to be a stripped-envelope event to be able to produce rapid core rotation that eventually gave birth to the magnetar. In context with SN Ib/c progenitor initial mass derived from our study, the two results are in line. More samples of both Ib/c SN and magnetar progenitors would be beneficial to establish a firm connection between the two. 

We note that this study is based on the assumption that the SN progenitor had been a member of the apparent host star cluster present at the explosion site. While this assumption might not necessarily be true, we argue that it is more likely that the associated cluster is the real parent stellar population of the SN progenitor rather than the diffuse background stellar population. The host cluster has been revealed to have very young age, thus the IMF implies higher probability of having high-mass stars compared to the background stellar population which is significantly older than the few Myr age of the prominent cluster. We discuss the probability of contamination by older, invisible clusters in the field in Paper II, where the effect is more important for type II SN progenitors, which are presumably older than SN Ib/c progenitors. 

We also present our mass estimate if we only consider the high-association cases only, i.e. SNe within 150 pc of the host cluster center. This 150 pc is about the intermediate size between classical and giant H\texttt{II} regions \citep{crowther13}, as most of our clusters are apparently of around this size. With this limitation, the mean progenitor mass of SNe Ic reduces to $27.5\pm6.7$~M$_\odot$, and SN Ib progenitors to $22.6\pm12.1$~M$_\odot$. This would not change the aforementioned conclusion.

\bigskip
\section{Summary}
\label{summ}

We present the results of our investigation of nearby SNe Ib/c explosion sites using integral field spectroscopy (IFS) technique, performed using UH88/SNIFS and Gemini/GMOS. Taking advantage of IFS, we spatially identified star clusters at the explosion sites and extracted their respective spectra for the purpose of analysis. From the spectra we derived the metallicities of the clusters in terms of 12+log(O/H) using the  strong line method, and age using comparison of observed H$\alpha$ emission equivalent width with theoretical SSP models from Starburst99. 

The method employed in this study offers the best available determination of SN progenitor natal metallicity, probing a very localized region specifically the parent stellar population from which the SN progenitor was born. This minimizes the effect of contamination from other populations and definitely is more reliable compared to metallicity determinations using proxies. As metallicity is one important parameter in context of SNe Ib/c progenitor studies, our result provides an important insight in this field. We found indication that the metallicity of SNe Ic explosion sites is on average higher than SNe Ib sites, by $\sim0.1$ dex. This result is in accordance with \citet{leloudas11} and \citet{sanders12} who also found indication that SN Ic explosion sites are more metal-rich compared to SN Ib sites, although not statistically significant, and \citet{modjaz11} who suggest the same conclusion with results they claim to be statistically significant.

The age of the SN progenitor host cluster is considered as the lifetime of the progenitor star. Comparing with Padova stellar evolution models, we are able to derive initial mass of SNe Ib/c progenitors from the ages of their parent star cluster. We found that SNe Ic explosion sites have younger age than SNe Ib, indicating that they arise from higher mass progenitors than SNe Ib. The derived progenitor mass from host cluster age is significantly higher for SN Ic, over 42~M$_\odot$ versus 21~M$_\odot$ for SN Ib progenitors. We note that these derived average progenitor mass were derived from a sample that includes sub-WR mass progenitors ($\lesssim25$~M$_\odot$) which may be interpreted as binary progenitors. If we consider the single progenitors only, SN Ic progenitors average 47.9~M$_\odot$ while SN Ib progenitor mass average is 32.2~M$_\odot$ -- still attesting that SN Ic progenitors are the more massive ones. Further, if only SNe showing high association with the parent clusters are considered, the average progenitor mass estimates would reduce to 27.5 and 22.6, for SNe Ic and Ib respectively.

In general, our findings support the current picture of stripped envelope (Ib/c) SN progenitors: SNe Ic tend to have more massive progenitors and more metal-rich than SNe Ib, and for both SN types interacting massive binary stars may comprise a fraction of the progenitors of the SNe.



\bigskip
\acknowledgments
We acknowledge the anonymous referee for helpful comments and suggestions.
H.K. acknowledges generous support from the Japanese government MEXT (Monbukagakusho) scholarship. Useful help from R. Pain, S. Rodney, and P. Weilbacher on working with datacubes is appreciated. 
We thank G. Leloudas for carefully reading the draft and providing important comments. We also thank J. Sollerman and F. Taddia for helpful comments on the draft of the manuscript.
We acknowledge excellent support from Gemini Observatory staffs for our observation.
This work is based on the data collected by using the Gemini Telescope, the opportunity for which was made available by making use of the inter-observatory time-exchange framework between Gemini and Subaru Observatories.
This work is based on the data by using the University of Hawaii 88-inch Telescope (UH88), for which the telescope time was afforded by the funding from National Astronomical Observatory of Japan.
This work was supported in part by a JSPS core-to core program "International Research Network for Dark Energy" and by JSPS research grants.
The work of K.M. is supported by World Premier International Research Center Initiative (WPI Initiative), MEXT, Japan, and Gant-in-aid for Scientific Research (23740141).
G.A. was supported by the Director, Office of Science, Office of High Energy Physics, of the U.S. Department of Energy under Contract No. DE-AC02-05CH11231. SNIFS on the UH 2.2-m telescope is part of the Nearby Supernova Factory II project, a scientific collaboration among the Centre de Recherche Astronomique de Lyon, Institut de Physique Nucl\'eaire de Lyon, Laboratoire de Physique Nucl\'eaire et des Hautes Energies, Lawrence Berkeley National Laboratory, Yale University, University of Bonn, Max Planck Institute for Astrophysics, Tsinghua Center for Astrophysics, and the Centre de Physique des Particules de Marseille.
This research has made use of the SIMBAD database and ALADIN, operated at CDS, Strasbourg, France.
This research has made use of the NASA/IPAC Extragalactic Database (NED) which is operated by the Jet Propulsion Laboratory, California Institute of Technology, under contract with the National Aeronautics and Space Administration.
Based on observations obtained at the Gemini Observatory, which is operated by the Association of Universities for Research in Astronomy, Inc., under a cooperative agreement with the NSF on behalf of the Gemini partnership: the National Science Foundation (United States), the Science and Technology Facilities Council (United Kingdom), the National Research Council (Canada), CONICYT (Chile), the Australian Research Council (Australia), Minist\'{e}rio da Ci\^{e}ncia, Tecnologia e Inova\c{c}\~{a}o (Brazil) and Ministerio de Ciencia, Tecnolog\'{i}a e Innovaci\'{o}n Productiva (Argentina). 
The authors wish to recognize and acknowledge the very significant cultural role and reverence that the summit of Mauna Kea has always had within the indigenous Hawaiian community.  We are most fortunate to have the opportunity to conduct observations from this mountain. 
\\



{\it Facilities:} \facility{UH88 (SNIFS, OPTIC)}, \facility{Gemini-N (GMOS)}.




\label{biblio}

\clearpage

\end{document}